\documentclass[onecolumn,pre,preprintnumbers,amsmath,amssymb]{revtex4}
\usepackage{graphicx}
\usepackage{dcolumn}
\usepackage{bm}
\input{epsf}

\begin{document}

\preprint{Scientometrics}

\title{ Organizational and dynamical aspects of  a small  network with two distinct communities : Neo creationists vs. Evolution Defenders}

\author{A. Garcia Cant\'u }
\email{agarciac@ulb.ac.be}
\affiliation{GRAPES,  SUPRATECS, Universit\'e de Li\`ege, B5 Sart-Tilman, B-4000 Li\`ege, Euroland
}
\author{M. Ausloos}
\email{marcel.ausloos@ulg.ac.be;  corresponding . author}
\affiliation{GRAPES,  SUPRATECS, Universit\'e de Li\`ege, B5 Sart-Tilman, B-4000 Li\`ege, Euroland
}

\date{24 April, 2008}

\begin{abstract}
Social impacts and degrees of organization inherent to opinion formation for interacting agents on networks present interesting questions of general interest from physics to sociology. We present a quantitative analysis of a case implying an evolving small size network, i.e. that inherent to the ongoing debate between modern creationists  (most are Intelligent Design (ID) proponents (IDP)) and Darwin's theory of Evolution  Defenders (DED)). This study is carried out by analyzing the structural properties of the citation network unfolded in the recent decades by publishing works belonging to members of the two communities. With the aim of capturing the dynamical aspects of the interaction between the IDP and DED groups, we focus on $two$ key quantities, namely, the  {\it degree of activity} of each group and the corresponding {\it degree of impact}  on the intellectual community at large. A representative measure of the former is provided by the {\it rate of production of publications} (RPP), whilst the latter can be assimilated to the{\it rate of increase in citations} (RIC). These quantities are determined, respectively, by the slope of the time
series obtained for the number of publications accumulated per year and by the slope of a similar time series obtained for the corresponding citations. The results indicate that in this case, the dynamics can be seen as geared by triggered or damped competition.  The network is a specific example of marked heterogeneity in exchange of information activity in and between the communities, particularly demonstrated  through the nodes having a high connectivity degree, i.e. opinion leaders.
\end{abstract}


\maketitle

\section{INTRODUCTION}

New areas of science continually evolve; others gain or lose importance, merge, or split. Progress can be made through some debate and argumentation based on experimental facts. It may happen that the (intrinsically theoretical) question at hands underlies exclusivity/monopolization principles. The more so when scientific aspects border philosophical ones. Let us disregard here the question of determinism or free will discussed from a quantum mechanics view point. Biological aspects like DarwinÕs evolution  (DE) theory raise many scientific (and other) questions. It has been challenged by intelligent design \cite{ID}  proponents (IDP). One neutral way to touch upon the problem is through statistical physics ideas, like recent studies of opinion formation on social networks. This subject of intense interest is considered here, from the point of view of a sociological network, - the DE $vs.$ ID debate is used as an illustrative set of data for $emerging$ hot topics.
 
Let us recall that among several topics of intense interest in modern statistical physics nowadays, networks \cite{randomnetwk,pastor,pastor2} and opinion formation \cite{weidlich,opinionf1,opinionformation} have led to a flurry of models, most of them assuming a large population of interacting agents, like the prey-predator models \cite{APCM04}. All these studies concern organizational processes of populations, on networks (or lattices) \cite{APE98,pnas.99.02.7821hub,PNAS-Netw-04slanina}.
These network systems are usually composed of a large number of internal components (the nodes and links), and describe a wide variety of systems of high intellectual and technological importance. Relevant questions pertain to the structure itself and the dynamics of properties, not mentioning the fundamental aspect of network evolutionary dynamics itself.   We will touch into such a problem observing how some structural aspect can indicate the presence of {\it opinion leaders} influencing therefore the group dynamics.

We consider that   a {\it spreading of information} topics, like ID and DE theory should be tackled along the lines of scientific investigation as done nowadays by physicists interested in social dynamics problems. We can consider that the actors are interacting agents, located at nodes of a network -to be characterized. Links are thereafter defined as citations. This is in line with studies on large-scale networks, like co-authorship networks \cite{newman,newman2b,newman2c,iina2}, namely networks where nodes represent scientists, and where a link is drawn between them if they co-authored a common paper.

 It is important to point that science (or more generally opinion) spreading is usually modeled by master equations with auto-catalytic processes \cite{andrea}, or by epidemic models on static networks \cite{holyst}. In this article, however, due to the limited data span and time scale, we propose to discuss the findings along the lines of a Verhulst population competition approach. 
Indeed away from any drastic transition the system does not  necessarily  evolve in a simple trivially smooth way.   Let us here recall that the most simple evolution of a biological population submitted to internal competition but in presence of a limited amount of resources was described by Verhulst \cite{verhulst} through the so called logistic map. The subject   has been quite vastly investigated, in particular through e.g. voter models mimicking social competition \cite{galam,sznajd} and others  obviously related to prey-predator  questions \cite{aperevPP,Pesceli}, population extinction risks \cite{aperisk}, etc. 

On the other hand let us stress that the identification of the mechanisms responsible for diffusion and, possibly leading to scientific avalanches  \cite{marsili} is primordial in order to understand the scientific response to external societal decisions, and to develop efficient policy recommendations: knowledge of hot topics, emergent research frontiers, or change of focus in certain areas is a critical component of resource allocation decisions in research laboratories,
governmental institutions, and corporations as pointed out in \cite{PNAS101.04.5287}.  Through the study of a bicommunity (necessarily small here) network we aim to put into evidence microscopic features and dynamics in a social network in an original way. Some other bipartite network study has also allow to describe heterogeneous communities \cite{uncovering}.

Among topics of opinion formation it is of interest to observe the evolution of subjects pertaining to the nature of how science is perceived or understood, either by scientists or by the public at large \cite{ballcriticalmass}. One such topics is creationism. No need to get involved in the pro and con \cite{con}, nor the reason why the subject is controversial, to say the least. Yet letters (to the editors), papers, media appearance \cite{hurd,miller} by true scientists or politicians or others are numerous on the subject for the last 15 years or so to provide data to be analyzed within modern lines. In Sec. II, we briefly recall the historical perspective in order to sustain our arguing for the method of investigation, outlined in Sec.III for constructing the network of interest.

 In order to build the network, we  first  identify the main factors of the Intelligent Design (ID) movement. On the basis of the article {\it Creationism and intelligent design} by R. T. Pennock \cite{Pennock},  which is criticizing ID, we can identify the founders of the ID movement. Moreover we use the ID web pages and their corresponding links. 
 Next we search how this predefined community is connected to defenders of the other, i.e. Darwin's evolution theory. We analyze the network structure, observing the emergence of opinion leaders through some novel mechanism, based on the so called notion of assortativity \cite{assor,RLMAtriangl}, in  Sec. IV. We use an empirical method to find out the main connecting nodes having links to both communities. Some discussion is found in Sec. V on the mutual interaction in and between these communities from a small network study point of view. Some conclusion is drawn in Sec. VI.

\section{Historical perspective}

One may distinguish two opinion groups according to the position of individuals about the subject of the origin of the universe and life. The first group holds the scientific consensus and in particular DarwinÕs evolution theory as a valid basis, whilst the second is formed by people adopting a theistic (rather biblical) view where natural processes are conceived as occurring out of the purposeful will of a supra natural entity. Amid them some belong to a historical movement, so called creationism, which aims to refute and overturn Darwin theory. Such organized opposition to evolution has found most of its adherents in different Christian traditions, actively engaged in promoting their values at the core of society. The influence of some of these religious groups has been especially relevant in some regions of the USA. Actually, at the beginning of the XX-th century several states in the American union approved laws $forbidding$  the teaching of evolution in public schools (see for  instance Ref.\cite{monkeytrial}); until the 1960s most of the school textbooks omitted evolution from their content. Furthermore, along the last two decades USA creationist organizations have
claimed the legitimacy of including their ideas in the biology teaching program of high schools, at the same level as DarwinÕs theory. These efforts to introduce creationism in the curricula of (public) schools led to some legal processes, none of which was successful in overcoming the USA constitutional principle of separation between Church and State.

 In this context, at the end of the 1980s the Intelligent Design Movement arises with the target of endowing the basic creationist ideas with the status of scientific status, or taking into account some bible sentences as reports (proofs even) of experimental facts. Thereby, the ID movement claims legitimacy on teaching the ensemble of their viewpoints as a scientific alternative to Darwin's explanation of the origin and diversity of life. Specifically, IDP have employed concepts of information theory, thermodynamics and molecular biology in seeking for evidences of an intelligent blueprint underlying the complexity observed in biological systems. Yet, none of the IDP arguments has been validated by most of the scientific community. Notice that not all IDP are Creationists, but the National Academy of Sciences and other similar groups have called intelligence design Ócreationism in disguiseÓ. In spite of this fact, the ID movement has further developed: since the second half part of the 1990s, in good measure thanks to the support and headquarter brought by the Discovery Institute's Center for Science and Culture (CSC) \cite{CSC}.  The ID idea has spread outside the USA, as can be seen in recent  publications in scientific journals. The increasing activity and impact of the ID movement has impelled the reaction of social and scientific organizations around the world. Among the most important ones, the non-profit organization National Center for Science Education (NCSE) \cite{NCSE} plays a relevant role in coordinating the activity of people defending the teaching of evolutionary biology in the USA. Hereafter we refer to the international group of people fighting ID as Darwin's theory of evolution defenders (DED).

As a matter of fact, understanding the ÓevolutionÓ followed by the IDP and DED groups requires accounting for their mutual interaction. This is an important element justifying our work.  As far as we know this aspect has remained unexplored, and is of scientific value whatever the pro and con arguments of these communities. The present work aims to contribute along this line by introducing a quantitative approach to the analysis of the dynamics of the interaction between both groups, as a study of a small network.

\section{Methodology}

With the aim of capturing the dynamical aspects of the interaction between the IDP and DED groups, we focus on $two$ key quantities, namely, the {\it degree of activity} of each group and the corresponding {\it degree of impact} on the academic  at large community. A representative measure of the former is provided by the {\it rate of production of publications} (RPP), whilst the latter can be assimilated to the {\it rate of increase in citations} (RIC). These quantities are determined, respectively, by the slope of the time series obtained for the number of publications accumulated per year and by the slope of a similar one obtained for the corresponding citations.

In order to gain insight on the degree of interrelation between the overall activity of any two social groups, it is useful to construct a network which structure represents the interactions between their constituting elements. For the present case, the most suitable network representation is one detailing the interactions established between individuals whenever they cite each others works. In particular we have constructed such citation network by applying the following procedure:

\begin{enumerate}

\item Search for nodes. Consider a short list $L_0$ containing the name of some of the ID proponents  (in particular we have chosen W. Dembski, M. Behe and S. Meyer). Using the Scholar Google Internet search tool, for each element of $L_0$ select a number of publications with a high rank of citations (for instance books) and create another list $L_1$ with  
 the different authors citing the elements of  $L_0$, while as objectively as possible recording their general positions upon either one  of the two sides of the debate. Next, introduce a set $N$ of nodes, each one corresponding to an element of $L_0$ $\cup$ $L_1$. In the present case, the data downloaded and examined  between Oct. 01 and Nov 15, 2007,  $N$ possesses 77 elements.
 \item Define the node state.   Introduce a partition in $N$ by endowing each node with an attribute (for instance to be in an up or down state) according to the apparent community position of its corresponding author.

 \item Insert links. For each pair of elements $ i,j$ in $N$, introduce a directed link $i \rightarrow j$ if, according to the outcome of the Scholar Google search process, the author associated to $i$ possesses at least one article citing another one written by $j$.
 
 \end{enumerate}
Thus we make a network representation of the IDP $vs.$ DED of the intra and inter relations of both competing groups. Beside  counting nodes and links we aim at observing some assortativity \cite{assor,RLMAtriangl} through triangular connexions.  We call {\it Undirected Link} (UL),  a link connecting a pair of nodes in both directions (A cites B and B cites A) and by extension a 
{\it Directed Triangle} (DT) the shortest cycle of a graph formed by ONLY directed links 
(A cites B cites C cites A) In this work we have not considered the {\it By-pass Triangles} (BT), i.e. the cases in which e.g. A cites B, B cites C and A cites C.

Once one has at hand a network representation of a social group, it is worth identifying the most influential individuals. In general, the most relevant elements of an interaction network can be identified by analyzing its local connectivity properties (see for instance Refs. \cite{newman,newman2b,newman2c,iina2}); this is called the $centrality$ of the node. Several techniques exist; one appealing technique bears upon the $Q$-method \cite{Flom,RousseauQ,ChenQ}, measuring the geodesic distance between nodes, but as the authors indicate can only be best applied by hand, on small networks, the algorithm being very complex to write. 
Yet, beside the fact that the communities have their own ''leaders'', obviously the most representative persons, an other interesting information would result from the  finding of the most relevant links connecting the communities. Since in the present work all links have the same weight, we can only rely on the node degree value to distinguish a node relative  importance. Since the interest resides in the interplay between communities,   here below we consider {\it a priori} that the centrality is quite well described and obtained if one looks at the number of inhomogeneous triangles to which a node belong. Thus, in the present case those individuals leading the opinion of   IDP and DED communities can be easily identified by analyzing the number of directed triangles and undirected links of the citation network. This fact allow us to obtain two pools containing, respectively, the most important opinion leaders of the IDP and DED groups.

\section{Results}

\begin{figure}
\begin{center}
\includegraphics[width=4.3in]{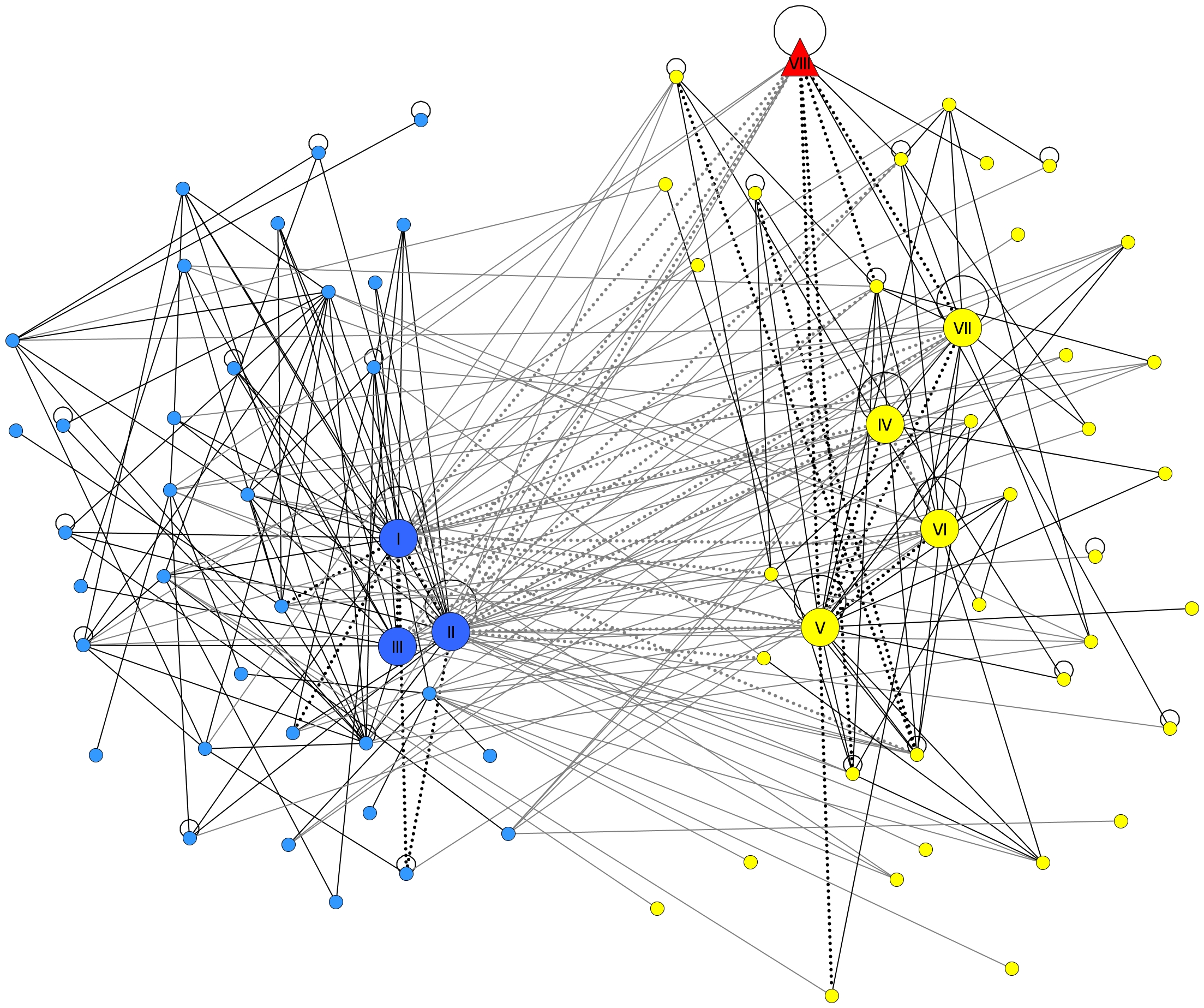}
\caption{\label{Network}
Schematic network representation of the interaction between elements of  IDP (dark (blue on line)
nodes; left hand side) and DED (light (yellow on line) nodes; right hand side) communities, according to data obtained from Google Scholar
database for the period 1990 - 2007.  Small loops indicate self-citations. Dotted lines correspond to so called  undirected links ; light  (dark) line indicate inter (intra) community links;
the largest circles represent the most relevant actors in both IDP and DED groups, i.e. (I) W.
Dembski, (II) M. Behe, (III) S. Meyer, (IV) K. Miller, (V) R.T. Pennock, (VI) B. Forrest
and (VII) E.C. Scott. The (VIII) (red)  triangle represents R. Dawkins (top right). The directionality of links is not indicated in order not to complicate the display
 }
\end{center}
\end{figure}

 Figure \ref{Network} exhibits the citation network obtained by applying the above procedure during the period October-November 2007. This network is composed by two subgraphs, one of 37 and the other one of 40 elements, corresponding to IDP and DED communities, respectively. There are 170, 128, and 217
 links in IDP, DED and intercommunity respectively, excluding self-citations.  The number of self citations can be found counting the number of loops in Fig.1;  it amounts to 11 and 15 respectively for IDP and DED.  Notice that we do not   at this stage give any weight nor directionality to the links.

  The results for the different possible types of triangles are given in Table I. We emphasize that triangles containing elements of different  communities are the most abundant ones.  Conversely, among the 348 triangles, 72 (thus 21 $\%$) are homogeneous relating only nodes of the same community.  Thus it is obvious that there is some interaction whence some dynamics. 
  
  Moreover the three main nodes in the ID community make up for 56$\%$ of the IDP triangles and 41 $\%$ of the inhomogeneous ones, while  5 nodes in the DE community make up for 51$\%$ of the inhomogeneous triangles (Table I).
Thus we can assume that a few so called opinion leaders can well describe the activity of   the whole group to which they belong. Therefore we restrict ourselves here below to construct the RPP and the RIC time series for both communities out of their corresponding pools of opinion leaders only. 
 
\begin{figure}
\hspace{-0.5cm}
\includegraphics[width=3.5in]{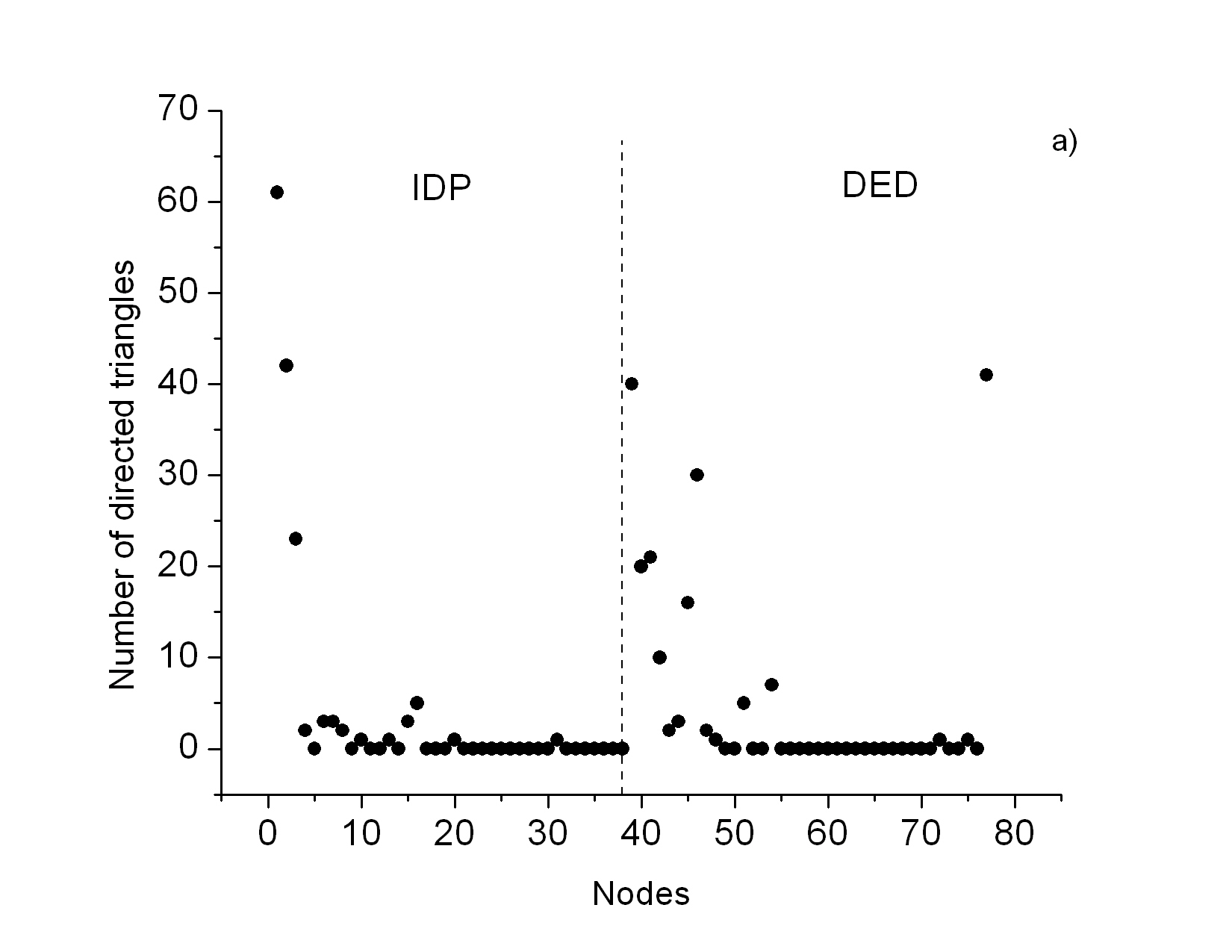}
\vspace{-0.4cm}
\includegraphics[width=3.5in]{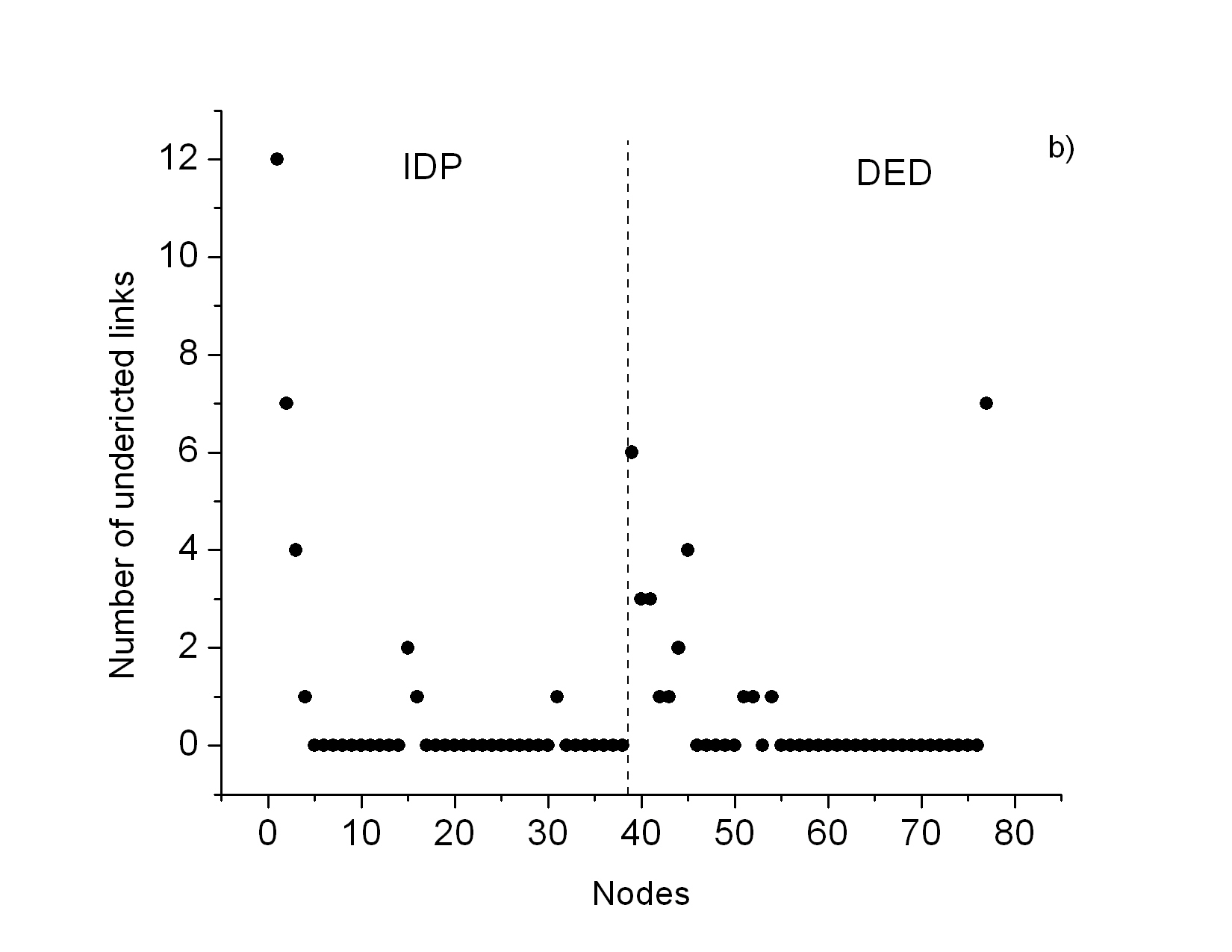}
\vspace{-0.5cm}
\caption{\label{CTUL} (a) Number of so called directed triangles; (b) Number of undirected links,  for the IDP and DED communities, plotted on left and right side of each figure respectively for nodes (authors) belonging to the interaction network displayed in Fig. 1 }
 
\end{figure}

 Let us   examine each node's $local$ context. Indeed not all authors are expected to belong to triangles. 
 Fig.  \ref{CTUL}  exhibits the number of directed triangles and (undirected) links for the different nodes of the citation network. These two plots show that the network nodes possess almost no triangle or undirected link, with the exception of a small set of nodes (authors) with very large number of both. Those well connected nodes correspond to {\it opinion leaders}, at the core of the discussion between ID and ED. Among them, for further study, we choose those nodes belonging to more than 20 triangles as the most influential ones, to be included in opinion leadership pools. According to this selection the most relevant elements of the IDP group are W. Dembski, M. Behe and St. Meyer. All these are founder fellows of the above mentioned  Center for Science and Culture \cite{CSC}. Both  Dembski and   Behe are authors of several books on ID  \cite{b1,b2,b3} and are the proponents of the  concepts of complex specified information \cite{nofreelunch} and irreducible complexity \cite{blackbox}, respectively. On the other hand, St. Meyer is a philosopher, director of the CSC and is the author of several writings promoting his ''God Hypothesis'' \cite{godhyp,godhyp2}. Most of the positions of St. Meyer are held by the ÒWedge ManifestoÓ, which is a document appeared on the internet in 2000 that lays down the guiding philosophy and the strategic plans of the ID movement. 

Concerning the DED,  this method of network analysis allows  us to identify R. Dawkins, R.T. Pennock, K. Miller, B. Forrest and E.C. Scott as the most important leaders of the group. Among them, R. Dawkins, a British evolutionary biology theoretician, author of several best-selling books popularizing a gene-centered view of evolution \cite{selfishgene},  is one of the most active critics of religion, whence of creationism. In particular, in his books ''The Blind Watchmaker'' \cite{blind} and "The God Delusion" \cite{delusion} Dawkins criticizes the idea of ID by developing an explanation on how the natural selection can explain the complex adaptability process of organisms. His argument has raised interest among the biology research community such that its impact extends beyond the context of the ID-Evolution discussion. Concerning the DE opinion leaders, E. C. Scott and B. Forrest are two of the most active directors of the NCSE. On the other hand, R.T. Pennock is a philosopher, associated member of NCSE and author of many books and articles critical of ID. His testimony as an expert witness was decisive for the United States federal court, in the so called Kitzmiller $vs.$ Dover Area School case \cite{kitzmiller,Lee,DeWolf}, to declare that ID is not science but another form of creationism. Finally, K. R. Miller is a (catholic) biology professor and author of articles and a book where he attacks creationism, while arguing that a belief in god and evolution are not mutually exclusive and has also been appearing in courts and the media. As mentioned above, the impact of works against creationism published by R.T. Pennock further extends beyond the context of the Creationism vs Evolution conflict. This is not the case of those written by  K. Miller, B. Forrest and E.C. Scott.  Those written by R. Dawkins have a stronger impact beyond the context of the creationism vs. evolution conflict. This fact has to be taken into account when analyzing the evolution of the interaction between IDP and ED groups. In the time series of annual publications and citations which we next present and analyze this is to be taken into account.

\begin{figure}
\centering
\includegraphics[height=7cm,width=7cm]{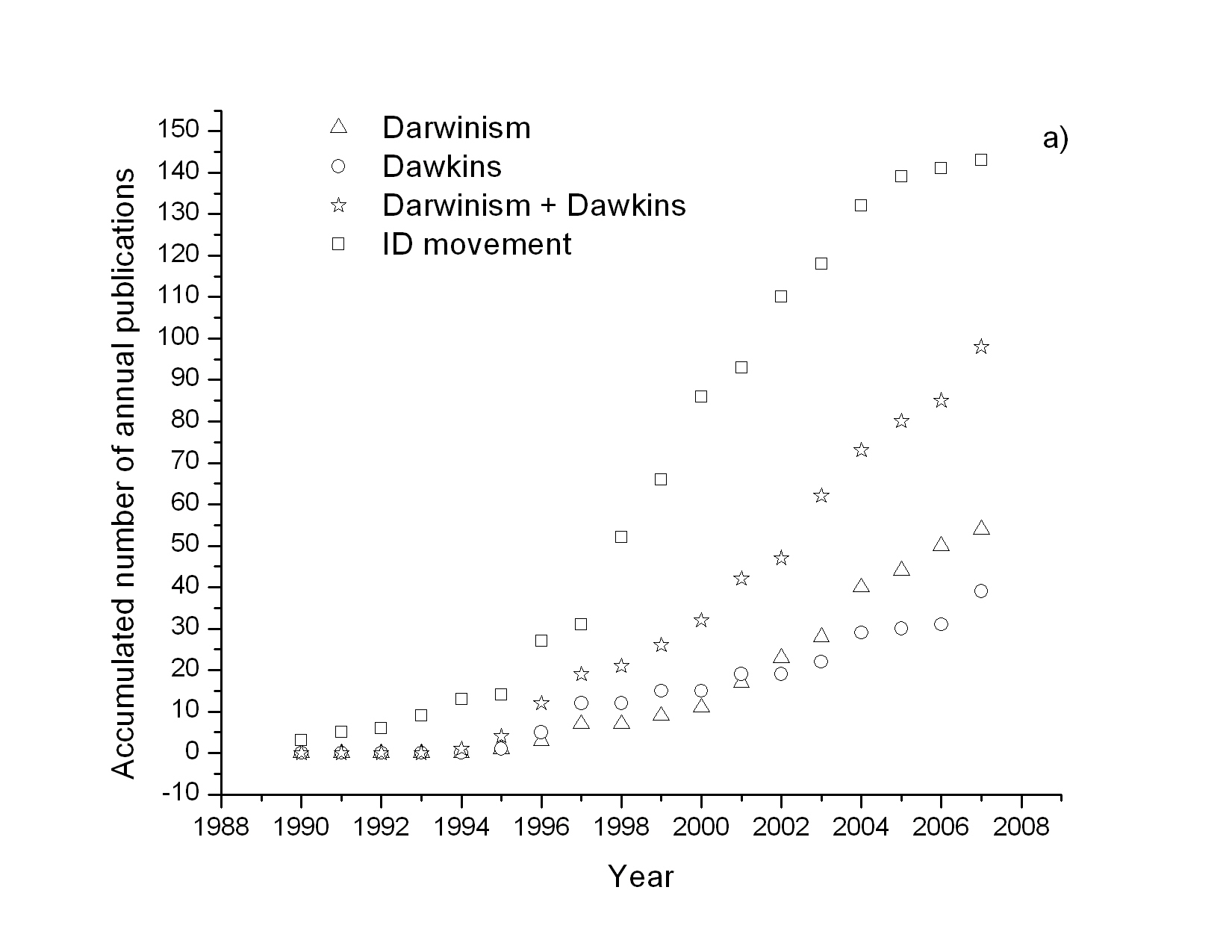}
\includegraphics[height=7cm,width=7cm]{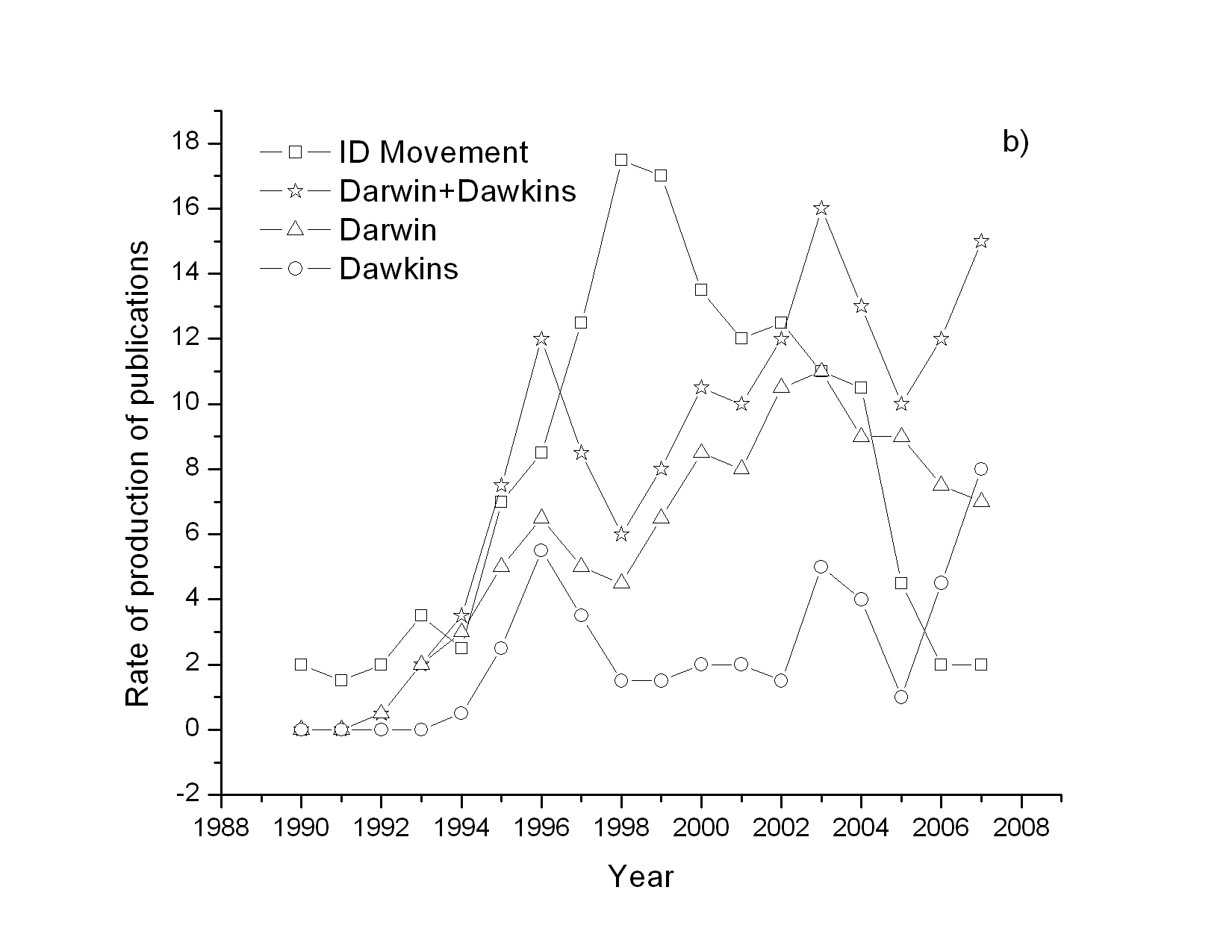}

\caption{\label{ANP}  Time dependence of the (a) accumulated number of publications (ANP); (b)  yearly  change in
production of publications, for authors being the nodes of the network shown in Fig. 1  $and$ being opinion leaders as defined in the main text}
 
\end{figure}

\begin{figure}
\begin{center}
\includegraphics[scale=0.3]{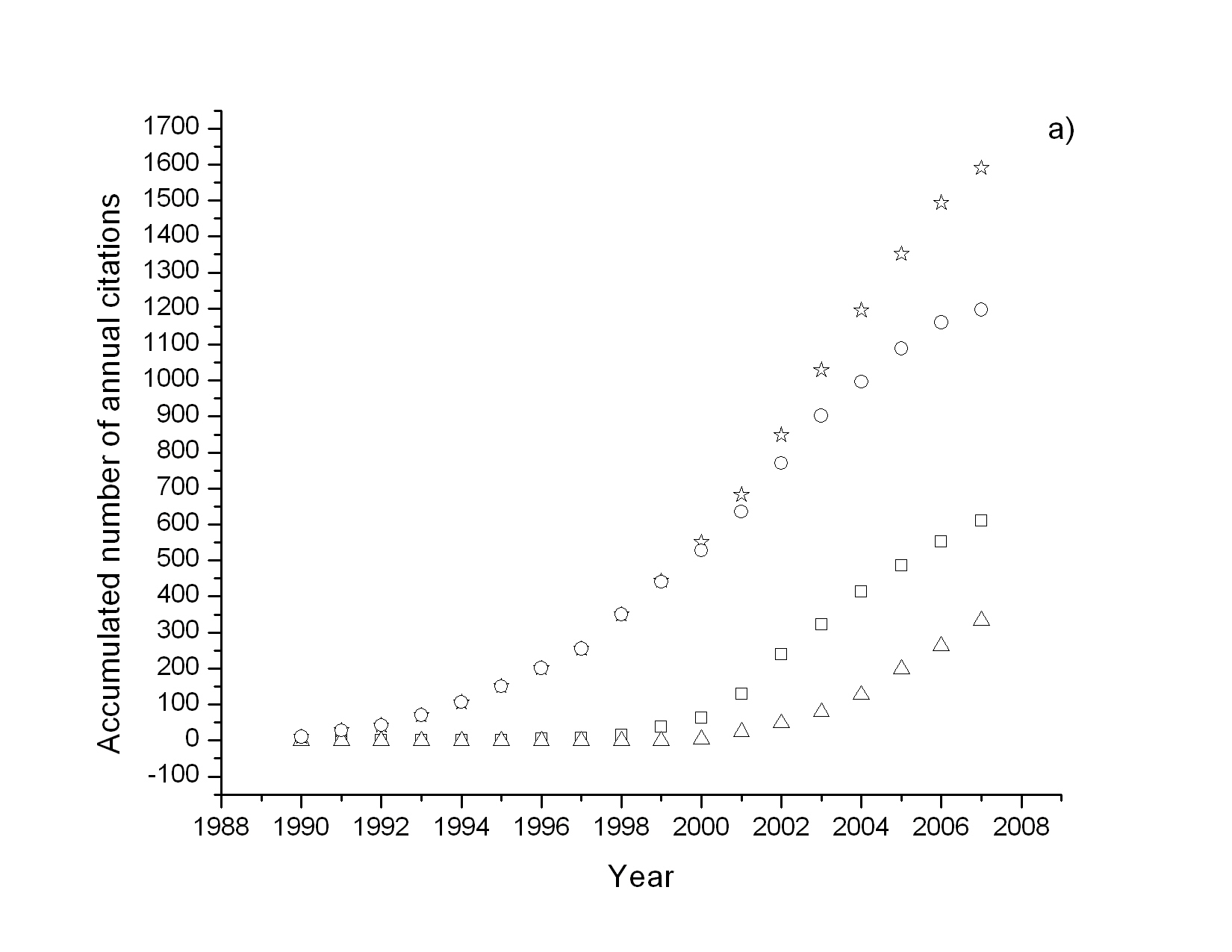}
\includegraphics[scale=0.3]{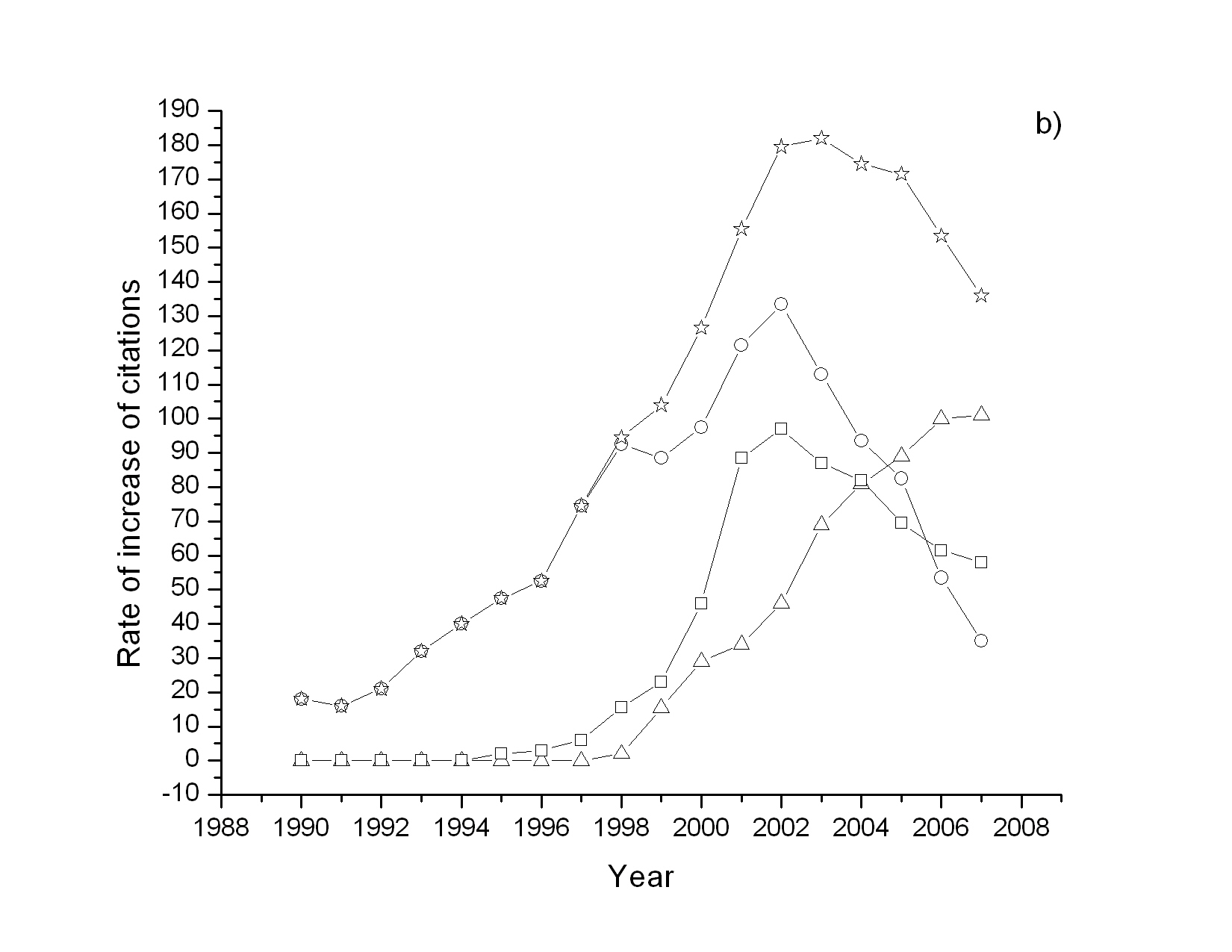}
\caption{\label{ANC} Time dependence of the (a) accumulated number of citations (ANC); (b)  yearly change  
 in number of citations, for authors being the nodes of the network shown in Fig. 1 $and$ being opinion leaders as defined in the main text ; same symbols as in Fig. 3}
\end{center}
\end{figure}

Let us display on
Fig. \ref{ANP}     the time series of the cumulative annual number of publications (ANP)    for the following group leaders : for ID (W. Dembski, M. Behe and St. Meyer), for ED (R.T. Pennock, K. Miller, B. Forrest and E.C. Scott) and, separately, R. Dawkins. We show also, in Fig. 3b, the  numerical derivative of  the data in Fig. 3a, i.e. the yearly rate of ANP, for authors being the nodes of the network shown in Fig.1.
Similarly let us display on
Fig. \ref{ANC}   the time dependence of   (a) the annual  number of citations (ANC) and  (b) the yearly rate of ANC, for the leaders of opinion, i.e. authors being on the nodes of the network shown in Fig. 1 but having more than 20 triangles. We do not distinguish between intra or inter community member citations at this level.

Finally, beside the leaders of opinion,  a referee suggested that we define which persons act as ''bridges'' between the two communities. In order to do so one could use Flom et al.'s  Q-measures  \cite{Flom,RousseauQ,ChenQ}. However there is no algorithm known for the calculation of so called geodesic measures. In fact the authors rely on  visual observation and counting. Thus we rely on reasoning to obtain such an information. The nodes being  the more frequent ''bridges'' between communities can be obtained from the inhomogeneous triangles, selecting the nodes  which have the more links between persons belonging to different communities. Observation of the network of Fig.1 and previously presented data indicate that it is better to check for those persons   who have the more links, thus are the opinion leaders  in the present case, i.e. those listed in Table II. More precisely it is found that  Behe and Dembski have  many links (26 and 25, respectively) with the DE community, Meyer coming a comfortable third with 15 links. It is remarkable that they have more links with the DE community than with their own ID community.  In contrast, all main DE nodes  have approximatively the same number  (each between 8 and 11) of links with the ID community but differ in link number with members of their DE community. 
 Pennock has quite many more intra links than other DE members.

 \begin{table}
\caption{Number of triangles  with specified edges between the three indicated nodes in the network of Fig.1}
\smallskip
\begin{footnotesize}
\begin{center}
\begin{tabular}{|c| c  c|  c c c  c|}
\hline 
Triangle configuration	&Number of triangles &
\\\hline   IDP-IDP-IDP 	& 21&
\\\hline   IDP-IDP-DED 	& 105&
\\\hline   IDP-DED-DED	& 171&
\\\hline   DED-DED-DED 	& 51&
\\\hline
\end{tabular}
\label{tab01} 
\end{center}
\end{footnotesize}
\end{table}

 \begin{table}
\caption{Most relevant nodes, i.e. opinion leaders,   in communities, according to the largest numbers of inhomogeneous triangles to which the nodes belong in the network of Fig.1. The number of undirected links, as defined in the text,  and respective intra and inter community links is shown for selected opinion leaders; "+s" means self-citation; in (out) means cited (citing) during the time interval so far considered}
\smallskip
\begin{footnotesize}
\begin{center}
\begin{tabular}{|c| c|  c|  c| c| c|  c| c|  }
\hline 
Names of & 	& Number of  & Number of & Number  of& Number of & Number of 
\\ node  &  Community & inhomogeneous   &   homogeneous  & undirected& intralinks  & interlinks 
\\ leaders &  &   triangles with &    triangles with  & links with & (in/out) &   (in/out)
\\\hline  W. Dembski 	&  ID & 55 &6& 12 &22 (14/8) +s &25 (17/8)
\\\hline   M. Behe &ID & 37 & 5& 7&17 (13/4) +s&26 (21/5)
\\\hline     S.C. Meyer &ID	& 20 &3& 4&11 (6/5) &15 (12/8)  \\\hline    
 & &&&& & \\\hline   R. Dawkins    & 	ED& 41 & 0&  7&13 (6/7) +s & 11 (7/4)
\\\hline   R. Pennock   & ED	& 40 & 0& 6&20 (11/9) +s &10 (4/6)
\\\hline    E.C. Scott          & ED	& 30 & 0& 4&12 (5/7) +s & 11 (5/6)
\\\hline    B. Forrest          & ED	& 21 & 0&3 &11 (7/4) +s  & 11 (2/9)
\\\hline   K. Miller    & ED	& 20 & 0&4 & 7 (4/3) +s & 8 (6/2)
\\\hline
\end{tabular}
\label{tab02} 
\end{center}
\end{footnotesize}
\end{table}

\section{Discussion}

The network has been constructed according to specified rules. We have obtained an approximately equivalent set of nodes belonging to two different communities. In view of the relative smallness of the network, it is irrelevant to observe whether scaling laws hold or not. Let us {\it in fine} consider the overall behavior of the bicommunity network from the question which is raised: whether there are interactions of relevance between the nodes and what dynamics can be observed. It is useful to reemphasize that the triangle study, related to the assortativity of the network, shows some  marked imbalance between homogeneous triangles, for which all nodes belong to a given community and those triangles having nodes belonging to different groups.
This  structural fact reflects the strong interaction between both opinion groups, from which therefore we can expect some non trivial dynamics.

In fact, we observe on Figs. 3-4 that the evolution of the number of publications  follow  a sigmoid curve remembering the logistic map. It is useful to distinguish that  Dawkins was  approximately the only DE author of interest, before 1997, but was joined by many other authors through some effect occurring in 1998 and accelerating later on. It can be observed on the rate of growth of publications  (Fig. 3b) that nevertheless all curves reached a maximum in 1996 and  2003, except the IDP.  The  yearly number of  publications by IDP has reached a maximum  in 1998, exactly when the other community has reached a minimum. This is an intuitively expected effect in competition games, and reminds us also of the behavior of enterprise competition on a market or of population competition in biological settings. It is interesting to observe that the minimum of IDP publications in  2005 coincides with a surge in publications for the other community. However it is hard to forecast at this stage whether some cycling effect will exist. We have not attempted a fit of the data  to some analytical form, because of the lack of meaning of the coefficients we could find and the intrinsically natural and large error bars (on both axes) known in such a scientometric analysis.

Nevertheless another aspect is also of interest, i.e. the relative  impact of the communities as well seen on Fig. 4. The rate of growth of the number of citations steadily increases but reaches a maximum for all except for the DED (less Dawkins) in  2001. The rate of growth of citations of the ID movement starts  in 1997 but a two year time lag occurs before the DED are quoted.  This value shows the interest of pursuing data analysis of bigger systems and  in the future. 

Finally let us recall that we have pointed out the relative numbers of inter- and intra-community links of the opinion leaders. The respective values are remarkably different.. Our purpose is not to  suggest psychological or philosophical causes, but merely to point out  here that from a network study point of view we have a specific example of marked heterogeneity in exchange of information activity in the communities and between them.

       \section{Conclusion}

Here above we have recalled that a central role in primordial science includes the emergence of Óscientific crisesÓ in inducing the appearance of major fields and trends in complex networks. There are inter-connections between distinct disciplines which induce a link, between author intra-connected otherwise through their discipline, i.e. communities on networks. This is also displayed in para scientific disciplines, and leads to the temptation of assimilating science to philosophy, metaphysics, religion. Undebatable (religious) faith is contradictory to the necessary doubtful attitude of true science, implying an apparent logical weakness of the latter. 

The creationism proposal and its subsequent sequel of propaganda publications is reminiscent of the diffusion of different topics in science and the creation of scientific avalanches, e.g. emergence of new research topics that rapidly attract large parts of the scientific community. We have not discussed the relevance, nor the importance of such different opinions, attempting to be as unbiased as possible at this level. Whence we have considered that it was of interesting value for examining this very modern case, as one made by two communities on a network. It was expected that there would be no phase transition like  phenomenon, one group influencing the state of the other  \cite{opinionformation}  through coherent fluctuations of opinion spreading. We have aimed at understanding some community evolution dynamics through the production of publications and through their impact, i.e. the rate of citations, intra- and inter-communities. 

In order to build the citation network, we have applied the usual method  \cite{newman2b,newman2c}, 
namely we have considered a network of authors placed at nodes, with a link between them if they cite  another's paper.  E.g.  we discriminated two  sub-communities:  one for scientists favoring {\em "Darwin's theory of evolution"}  and others proposing {\em "creationism"} [as a scientific alternative ]. Moreover, we have considered  a specific scientist as an opinion leader, who have more than (arbitrarily) 20 (about 25$\% $)   connections with authors in his/her field of interest. These opinion leaders emerge as well known scientists.   They are also the "bridges" between the communities due to their much more important number of links than other members.

 It is relevant to recall that science spreading is usually modeled by master equations with auto-catalytic processes \cite{andrea}, or by epidemic models on static networks \cite{holyst}. In this article, however, due to the limited data span and time scale, we have proposed to discuss the findings along the lines of a Verhulst population competition approach.  It is intuitively obvious that there is no thermodynamic like transition to be expected since the communities are pretty much behaving as in a very deep potential well. There is not much change/fluctuations in the node state, whatever the link weights or number.  We have observed that the competition between the communities, exemplified by the leaders of opinions resemble a population competition as in other biological or financial settings.

{\bf Acknowledgements}
The authors would like to thank  A. Fronczak, I. Hellsten,  J. Holyst, R. Lambiotte, A. Scharnhorst, D. Stauffer and K. Suchecki for fruitful discussions.
This work  has been supported by European Commission {\it Critical Events in Evolving 
Network} (CREEN) project  FP6-2003-NEST Path-012864.
 The network in Fig.1 is plotted thanks to the {\em visone} graphical tool: {\em http://www.visone.de/}.

\end{document}